# Study and improvements of a radially coupled coaxial Fast Faraday cup design towards lower intensity beams

K. Mal[1], S. Kumar[1], G.Rodrigues[1], R. Singh[2]

[1]Inter University Accelerator Centre, New Delhi 110067, India

[2]GSI Helmholtzzentrum für Schwerionenforschung GmbH, Planckstraβe 1, Darmstadt, Germany

**Abstract**

A radially-coupled coaxial fast Faraday cup design was presented in [1] for high-intensity non-relativistic proton beams. In this work, we discuss a modification of that design in the context of a relatively lower intensity ion beam for longitudinal charge profile measurements. Particle-in-cell and time domain electromagnetic simulations of the new design with a focus on avoiding field dilution while generating enough signal for relatively lower intensity ion beams at the upcoming High Current Injector Programme at IUAC, New Delhi [2,3] is discussed. Profile distortions from secondary electron emission are estimated and strategies to suppress them are discussed.

# Introduction

The operation of Faraday cups for intensity measurements is well established where the suppression of the emitted secondary electrons is carried out using a superimposed electric field, such that the emitted secondary electrons are retarded and recaptured [4] . For charge profile measurements of longitudinal short bunches ($\leq$ 5 ns), it is critical to avoid impedance discontinuities in the Faraday cup structure until frequencies upto few GHz. Modified Faraday cup designs tailored to measure longitudinal charge distributions [4,7] are called Fast Faraday Cups (FFC). Early FFC designs were tapered extension of coaxial cables allowing for full beam deposition on the central conductor while maintaining $50\,\Omega$ characteristic impedance [4,5]. Following that, alternative FFC designs based on radial coupling in the central conductor [1] of a co-axial cable and microstrip based designs [6,7] have been used in various accelerator laboratories. So far, most studies available in the literature on FFC are focused mainly on electromagnetic characteristics of the FFC, i.e. targeting impedance mismatch aspects. However, additional challenges for short bunched beam measurements in non-relativistic regimes are a) the field elongation and b) distortion by the emission of secondary electrons. In this study, we will present an adaptation of a Radially-Coupled Coaxial Fast Faraday Cup (RCFFC) [1] for beam conditions available at the High Current Injector (HCI) Programme, which is presently under commissioning stages at the Inter-University Accelerator Center(IUAC),New Delhi [2,3]. The primary drawback of the original design [1] is the low 'signal-to-noise' ratio due to the narrow beam limiting aperture of $0.8\,\text{mm}$ along with the need for precise beam alignment with long averaging times for the measurement. For larger 'signal-to-noise' ratio, we increased the aperture size and used a transition method [8] without curved structures for the transition from the N-type connector to the cup region to achieve uniform impedance and low reflection. Second challenge is the delayed signal induction due to emission of secondary electrons. The major modifications in our adapted design to counter these aforementioned challenges are discussed in this contribution. We also herewith discuss signal induction process, field dilution, secondary electron emission aspects, fabrication challenges as well as thermal considerations on the simulated design.

## EM simulations of a modified RCFFC

The characteristic impedance ($Z_{coax}$) of a coaxial transmission line is defined as follows [9] :

$$Z_{coax} = \frac{1}{2\pi}\sqrt{\frac{\mu}{\epsilon}}\ln\left(\frac{r_2}{r_1}\right) \quad (1)$$

where and are permittivity and permeability of the material, respectively, $r_1$ and $r_2$ are the radii of the inner and outer conductor, respectively. It is important to note that a coaxial line is the upper-frequency limit for pure Transverse Electro-Magnetic (TEM) mode operation and is referred to the frequency at which the first non-TEM mode starts to propagate. The non-TEM mode with the lowest cut-off frequency ($f_c$) is $TE_{11}$ given by the following relation [9]

$$f_c = \frac{c}{\pi\sqrt{\mu_r\epsilon_r}(r_1+r_2)} \quad (2)$$

where $\mu_r$ and $\epsilon_r$ are the relative permeability and permittivity of the medium, respectively and c is the speed of light. Therefore the width of the co-axial line cannot be arbitrarily increased. To counter this aspect, a conical taper between a thin and a thick co-axial line is chosen. The characteristic impedance ($Z_{con}$) of a conical line is defined as follows [10]:

$$Z_{con} = \frac{1}{2\pi}\sqrt{\frac{\mu}{\epsilon}}\ln\left(\frac{\tan(\theta_2/2)}{\tan(\theta_1/2)}\right) \quad (3)$$

where $\theta_1$ and $\theta_2$ are the angles formed by the inner and outer conductor of the conical line with the symmetry axis (z-axis) (See Figure 2.). In order to minimise the reflection at the transition of coaxial and conical lines, it is necessary to design in such a way that the characteristic impedances of both the lines have the same impedance of 50 Ω. However, equation 1 and 2 can never be equal except when $\theta_1$ and $\theta_2$ are identical and can asymptotically approach each other for long lines such that $\theta_1$ and $\theta_2$ approach zero [10]. The electromagnetic simulation code, CST Microwave Studio [11] was used to design the RCFFC assembly with the interfacing option to the transmission line at the desired characteristic impedance of 50 Ω. Figure 1. shows the cross-sectional view of the CST simulation model of the RCFFC.

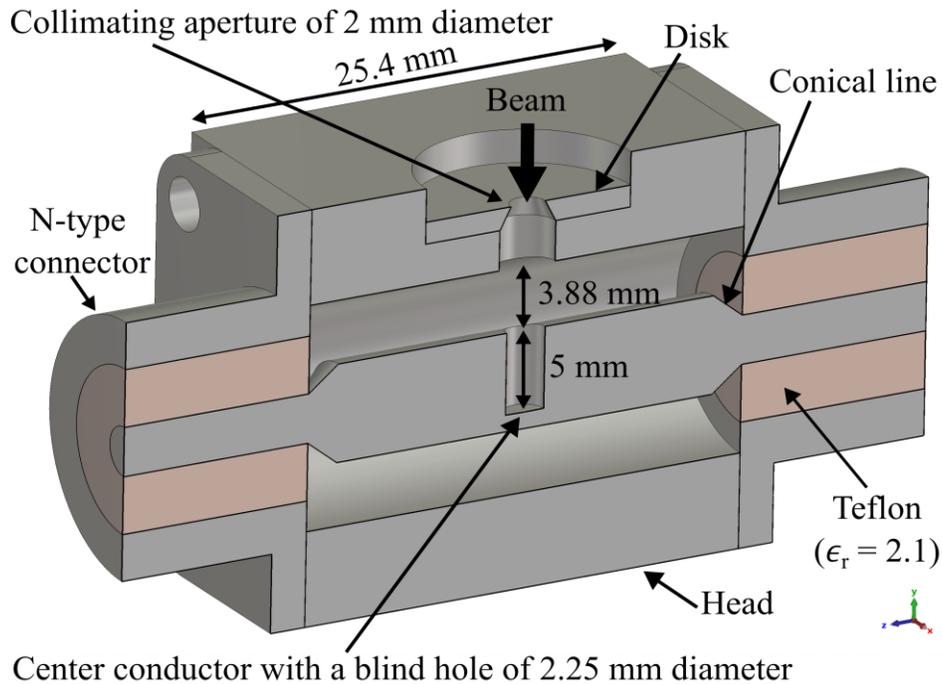

Figure 1. CST simulation model (cross-sectional view) of the RCFFC with the geometrical parameters.

The RCFFC consists of a metallic cube with two N-type connectors positioned concentrically on its side as shown in figure 1. Note that the collimating aperture of diameter 2 mm is chosen based on the relatively lower beam intensities available at the High Current Injector. A rod of diameter 6 mm was inserted between the central electrodes of the connectors and tapered down to 3 mm, which is the diameter of the standard N-type connector pin. The rod acts as a collector as well as the transmission line's inner electrode. The outer conductor of the conical line is bent at $\theta_2 = 90°$. The diameter of the collector hole was chosen to be 2.25 mm to avoid hitting the beam directly. The secondary and reflected particles stay mostly inside the collector hole because the depth of the hole is chosen at 5 mm, which is twice as large as its diameter. With these chosen dimensions, the first-order estimate of the RCFFC geometry parameters has been calculated using the analytical formula (equation 1) and (equation 3), and are shown in figure 2(a).

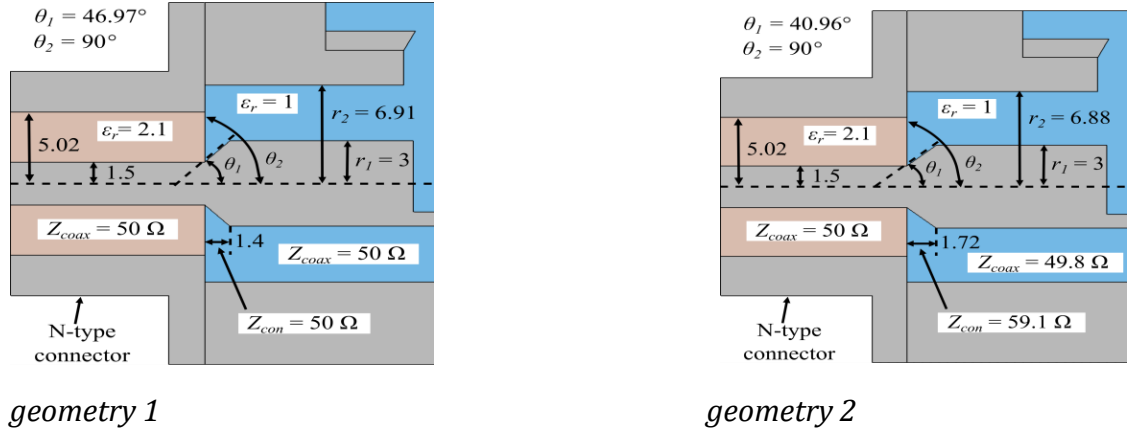

geometry 1                                      geometry 2

Figure 2. Left side of the RCFFC geometry and its design parameters (unit: mm) for characteristic impedance of 50W, (a) calculated using analytical formulas and (b) optimized using the CST-MWS.

For the chosen dimensions and the medium properties, as shown in figure 2, the cut-off frequencies of the N-type connector and RCFFC head are 10.10 GHz and 9.64 GHz, respectively. In order to obtain the return loss ($S_{11}$) and insertion loss ($S_{21}$), a two-port analysis [12] was carried out by assigning two waveguide ports at the left and right N-type connectors, respectively. In addition, a Time Domain Reflection (TDR) analysis [13] was also performed to compute the distributed characteristic impedances along the coaxial and conical lines of the RCFFC. Using the parameters as described in figure 2(a). The EM wave simulation was performed by considering only TEM mode at the waveguide ports with a frequency range up to 9 GHz to evaluate the performance of the RCFFC. Figure 3. shows the simulated return loss at the input port, insertion loss between two ports and the characteristic impedance ($Z_c$) along the conical and coaxial lines for the non optimized geometry as shown in figure 2 (a). It can be observed that the return loss of the RCFFC depicts two resonance peaks below the cut-off frequency (10.10 GHz) which are due to impedance mismatch at the conical lines and the center of the RCFFC, although, the insertion loss is not so large. Similarly, it can be also seen that the characteristic impedance of the RCFFC is almost equal to theoretical impedance (50 Ω) but it also shows two capacitive peaks at the conical line and one inductive peak at the center of the RCFFC. The difference between the theoretical and simulated impedance at the conical line could possibly be due to the short length of the conical line. The inductive peak at the center of the RCFFC is due to the holes in the collector electrode and in the head of the RCFFC.

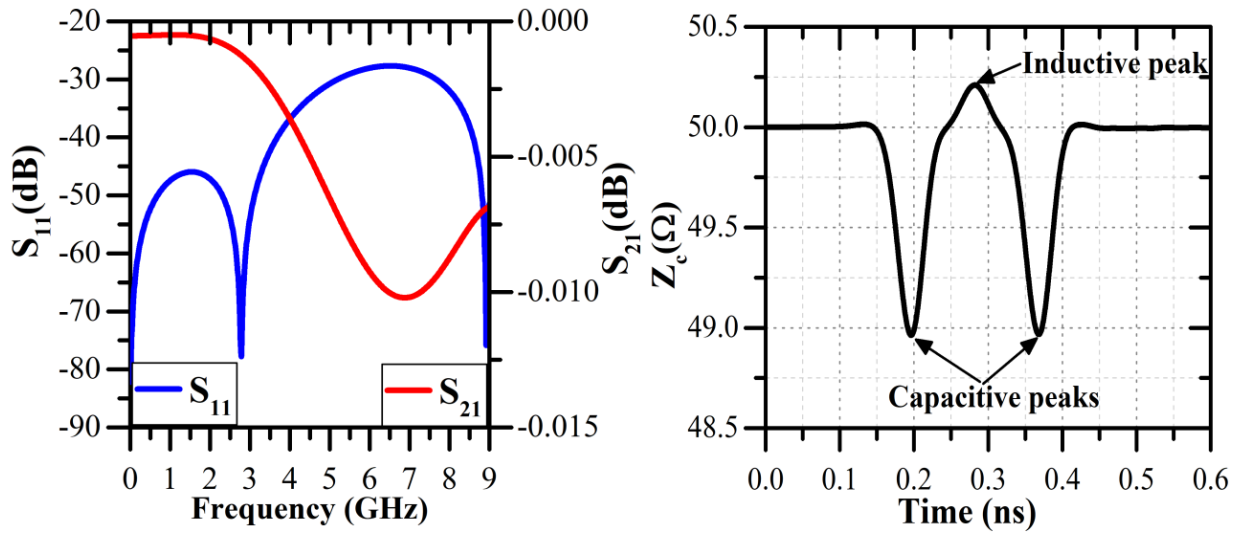

Figure 3. CST Simulated results of geometry 1 (a) return loss at the input port, and insertion loss between two ports, (b) characteristic impedance (Zc).

To achieve uniform impedance and minimize the reflections at the junction between the conical and coaxial lines, the length of the conical line and inner radius of the outer conductor ($r_2$) were optimized. The optimized geometry parameters are shown in figure 2(b). For the optimized geometry, the length of the conical line is 1.72 mm and thus, the angle ($\theta_1$) of the inner conductor of the conical line turns out to be 40.96°. The optimized inner radius ($r_2$) of the outer conductor is

6.88 mm to suppress the inductive peak at the center of the RCFFC. The theoretical characteristic impedance of the conical line and head of the RCFFC are 59.1 Ω and 49.8 Ω, respectively. The simulated return loss, insertion loss and characteristic impedance of the optimized geometry are shown in figure 4.

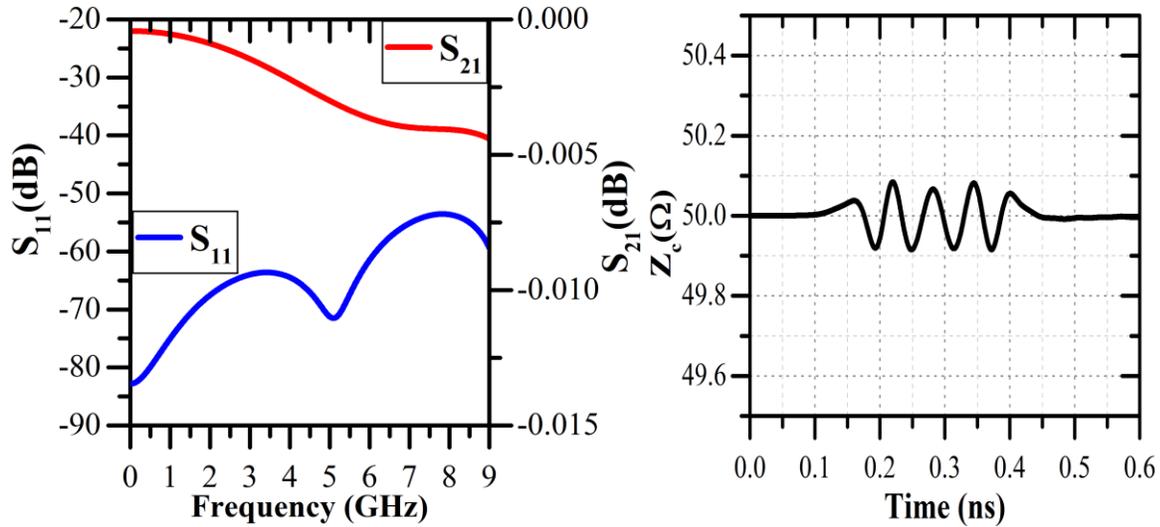

Figure 4. CST Simulated results of geometry 2 (a) return loss at the input port, and insertion loss between two ports, (b) characteristic impedance (Zc).

It can be seen from figure 3(a) and 4(a), that the return loss is reduced from −30 dB to −55 dB. The insertion loss for the optimized geometry is also reduced. The impedance difference between the conical and coaxial lines was also significantly reduced and is almost equal to 50 Ω (see figure 4(b)). We can see that the insertion loss of the optimized RCFFC geometry attains a value of −0.005 dB at frequency 9 GHz. Using the well-known relation between the bandwidth (BW) and signal rise time (t) (BW = 0.35/t), the rise time of the RCFFC geometry can be improved to a value, better than 39 ps.

The electric and magnetic field probes are introduced in three places: N-type connector, conical line, and near center of the RCFFC. The wave impedance ($Z_w$) which is the ratio of the electric and magnetic field is calculated at these three locations in the geometry 1 and geometry 2 (optimized).

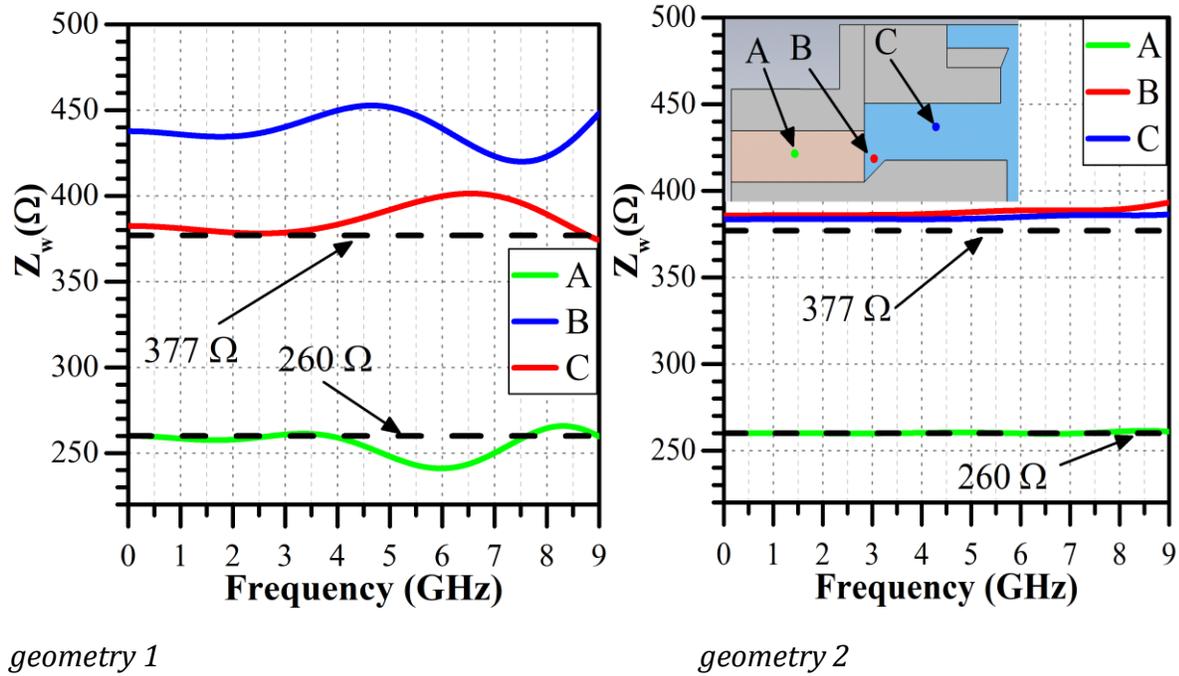

*geometry 1*                    *geometry 2*

Figure 5. Comparison of wave impedance ($Z_w$) at three locations.

As we can see that from figure 5(b) that the wave impedance at the conical line and near the center of the RCFFC for the geometry 2 are almost uniform and equal to 377 Ω which is the wave impedance of the TEM-wave in the vacuum. Similarly, the wave impedance at the point A in the N-type connector is also uniform and equal to 260 Ω, which is the wave impedance of the TEM-wave in teflon medium ($\epsilon_r = 2.1$). Looking at the wave impedance at three locations in geometry 1 as shown in figure 5(a), the resonant ripples due to impedance mismatching at the conical line appear and are larger than that of geometry 2 as the frequency increases. It can be also seen that the wave impedance at the conical line is higher than 377 Ω. This may be due to the low characteristic impedance of the conical line (see figure 3(b)). The high capacitance indicates a strong electric field, which results in a high wave impedance. As a result, when using characteristic impedance larger than 50 Ω at the conical line, it shows a more uniform and excellent matching characteristics (shown in figure 2(b) and 5(b)).

**Robustness of the design parameters**

The effect of manufacturing tolerance of various parameters on the characteristic impedance of the RCFFC is investigated. The diameter of the collector electrode ($d = 2r_1$), inner diameter of the outer conductor ($D = 2r_2$) and the transition length ($L_{cone}$) of coaxial to conical were increased by 0.1 mm individually about their optimized values (see figure 2(b)). Figure 6(a) and 6(b) shows the simulated reflection coefficient and the characteristic impedance of the RCFFC, respectively when the diameter of the collector electrode, inner diameter of the outer conductor, and the transition length of the coaxial to conical line were increased by 0.1 mm. For the comparison, the reflection coefficient and the characteristic impedance for the optimal parameter values are also included in figure 6. We can see that a change of 0.1 mm in the dimensions of the RCFFC causes significant changes in the reflection coefficient and characteristic impedance, specially for the collector electrode (see figure 6.). This tolerance effect must be carefully considered during fabrication of the collector electrode.

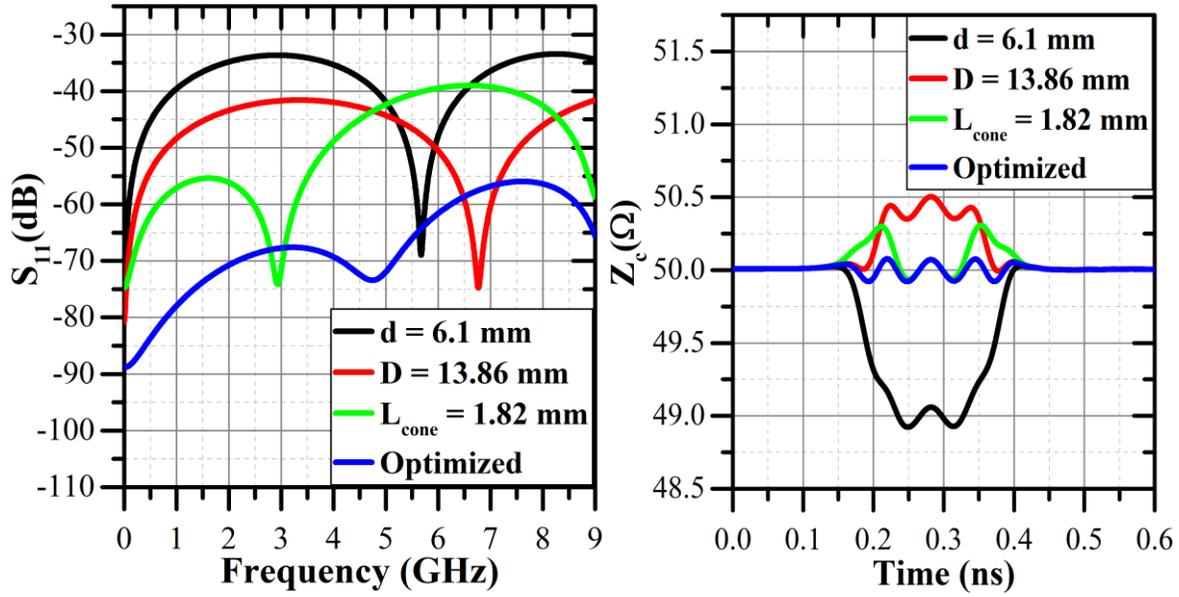

Figure 6. Simulated results showing the effect of individual tolerance on (a) Reflection coefficient and (b) Characteristic impedance of the RCFFC, when parameters are increased by 0:1mm about their optimal values (d = 6mm, D = 13:77mm, and L$_{cone}$ = 1:72 mm).

Considering the worst case, where all the dimensions of the RCFFC are 0.1 mm off from their optimized values means $d = 5.9$ mm, $D = 13.8$ mm, and $L_{cone} = 1.82$ mm. Figure 7. shows the simulated reflection coefficient and characteristic impedance for the worst case, for the case when diameter of the collector electrode changed by 0.1 mm alone and for the optimal parameter values. From figure 7(a), we can see that a small change in the reflection coefficient was observed for the worst case compare to the case when the diameter of the collector electrode was changed by 0.1 mm alone. This further shows that how tight manufacturing tolerance is required for the collector electrode. Therefore, a margin of ±0.1 mm is left to account for the manufacturing tolerances in order to still obtain the reflection coefficient equal to −30 dB.

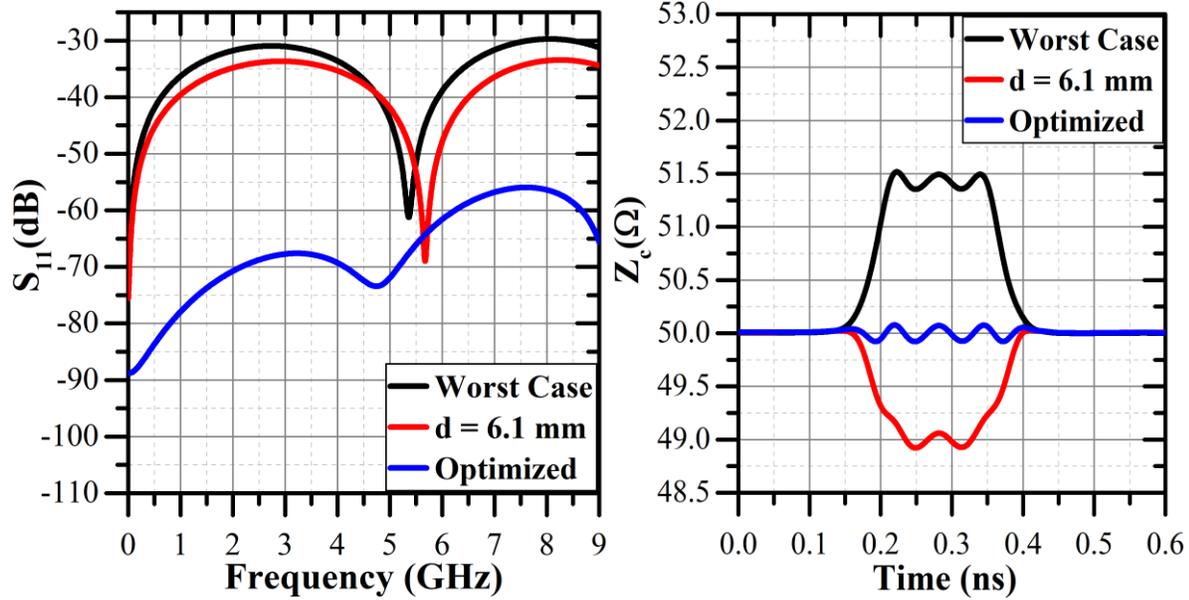

Figure 7. Simulated results showing the effect of tolerance on (a) Reflection coefficient and (b) Characteristic impedance of the RCFFC, for the worst case, for the case when d = 6:1mm, and for the optimized parameters (d = 6mm, D = 13:77mm, and $L_{cone}$ = 1:72 mm).

The optimized design of RCFFC as shown in figure 2(b), although showing excellent performance but requires tight manufacturing tolerance specially for the diameter (6 mm) of the collector electrode.

**Emission of Secondary Electrons**

Apart from the characteristic impedance, some other properties should be considered in design of the FFC for measuring accurate bunch widths. The FFC is normally designed to stop a beam of charged particles (like protons, and heavy ions etc) and minimize the secondary electrons (SE) ejected from the collector when the projectile particles strike on it. To recapture all of the SEs, electrostatic suppression is used in many conventional FCs [14-16] and FFCs [5,17]. In our design, the electrostatic suppressor cannot be trivially implemented in front of the collector electrode, and to minimize the SEs, the structure blind hole in the collector was modified.

In order to minimize the loss of the SEs, their production mechanism, energy and angular distribution as well as the total yield (as a function of ion energy) are required. The charged particles interact mainly through the coulomb forces between positive charges of the projectile and orbital electrons of the target [18]. When an energetic charged particles hits a surface, secondary electrons are emitted. These SEs are produced by two types of collision processes: distant and close collisions. In case of distant collisions, a small energy transfer take place which give rise to a large number of low-energy secondary electrons. In case of close collisions, a large energy transfer take place which result in excitation of a small number of energetic $\delta$-electrons which may produced further electrons by cascade processes [19]. The energy distribution of the secondary electrons has a peak at few eV with a full width at half maximum (FWHM) of the same order of magnitude, thus about 80-95% of the ejected electrons are below 50 eV [19]. The angular distribution of the secondary electrons follows the cosine law [21] and the total secondary electron yield ($Y$) as a function of angle of incidence is given by the following equation [19] ;

$$Y(\theta) = Y(0)\cos^{-1}\theta \qquad (4)$$

where $\theta$ is the angle of incidence of the projectile respect to the surface normal and $Y(0)$ is the total secondary electron yield at the normal angle of incidence. This is an empirical formula which valid for angles $< 70°$. Thus, increasing the angle of incidence or employing a cone shape at the bottom of the blind hole in the collector electrode, increases the geometrical path length of impinging particles within the secondary electron escape zone by a factor of $\cos^{-1}\theta$. As a result, the yield ratio drops to 0.866 for a 30° cone angle, 0.707 for a 45° cone angle, and 0.505 for a 60° cone angle.

For some targets materials, the total electron yield $Y$, as a function of proton energy, has been determined experimentally [22-26]. Haque et al.[27], proposed a model for the calculation of proton-induced secondary electron emission from elemental solid targets which is valid for a wide range of proton energy (1 keV to 1000 MeV) and shows a reasonable agreement with the available experimental data. As an example the value of $Y$ for proton energy 1.8 MeV hitting copper target is around 0.67 electrons per proton [27]. The beam intensity measurements will be highly uncertain if these electrons escape from the RCFFC. Therefore, it is required to suppress all of the back-scattered secondary electrons. Four different types of the back-scattered loss reduction techniques are quoted in reference [15] and we have used geometrical-based technique to minimize the loss of the back-scattered secondary electrons.

In order to stop the projectile ions of total energy 1.8 MeV/A, the thickness of the beam stopper should be much larger than the range of the projectiles. The Copper material is chosen for the collector of the RCFFC because of its good electrical property and its excellent thermal conductivity. The range of $^{14}N$ ions of energy 25.2 MeV in copper has been calculated to be 8.2 µm using SRIM code [28] and shown in figure 8(a). Similarly, the range of $^{238}U$ of energy 428.4 MeV in copper is to be 12.7 µm and shown in figure 8(b). Similar results have been also confirmed using online calculator [29]. Thus, choosing a thickness of 1 mm is sufficient to stop all ions of energy 1.8 MeV/A.

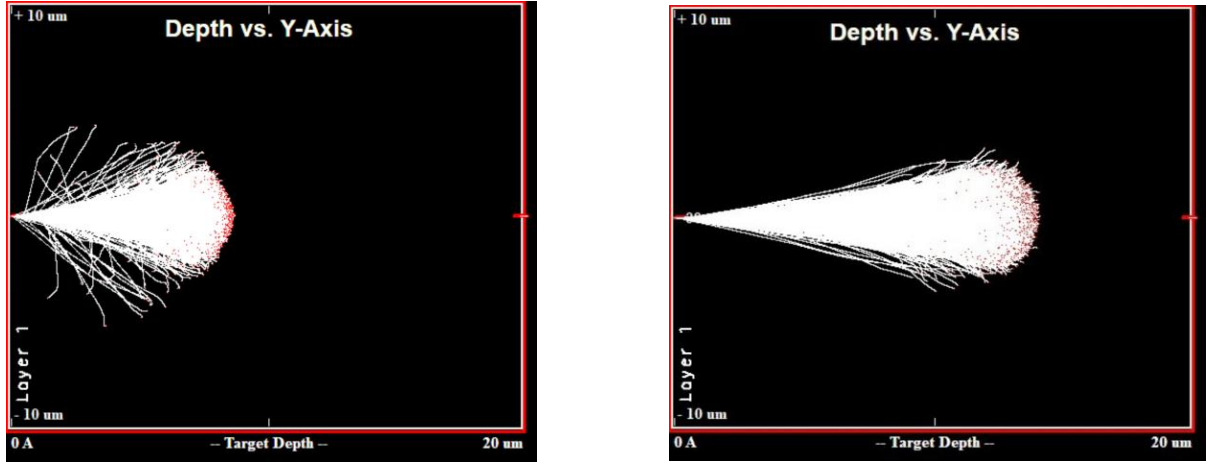

Figure 8. Range of ions in copper target of thickness 20um bombarded by (a) Nitrogen ions of energy 25.2MeV, and (b) Uranium ions of energy 428.4MeV.

In order to minimize the back scattered secondary electrons, the depth of the blind was varied (maximum 5 mm) and calculate the escape probability of SE using particle tracking solver of the Microwave CST Studio [11]. In Microwave CST Studio, the yield of the secondary electrons induced by the ions is given by the following [11];

$$Y(\theta) = Y(0)\left(1 + \frac{k_s \theta^2}{2\pi}\right) \quad (5)$$

where $\theta$ is the angle of incidence of the projectile respect to the surface normal. $Y(0)$ is the total secondary electron yield (SEY) at the normal angle of incidence and used 1.0 electrons/ion due to lack of available experimental data for heavy ions of energy 1.8 MeV/A. The parameter $k_s$ is a so-called smoothness parameter and depends on the surface of the material and We use $k_s = 1$ in the simulation. The energy spectrum of the secondary electrons is gamma distributed and given as follows [11] :

$$f(E) = Y(E_0, \theta_0)\frac{E}{T^2}\exp\left(\frac{-E}{T}\right)P^{-1}\left(2, \frac{E_0}{T}\right) \quad (6)$$

where $E$ and $E_0$ are the energy of the SE and the projectile ions, respectively. $Y(E_0, \theta_0)$ is the total SEY at projectile energy $E_0$ and angle of incidence $\theta_0$. T is the most probable energy (MPE) in eV and can be used to adjust the electron energy distribution function in equation 6 to different materials. P is the incomplete gamma function. In this simulation, we use $Y(E_0, \theta_0) = 1.0$ electrons/ion for nitrogen beam of energy 25.2 MeV at normal angle of incidence and $MPE = 7.5$ eV (default value for copper material).

To calculate the escape probability of the SE as function of depth of the blind hole, the following assumptions made the analysis simpler.

1. The trajectories of 10 k nitrogen ions of total 1 µA current and with energy 25.2 MeV were simulated. The nitrogen beam is assumed to be emitted from a circle of diameter less than the aperture size (2 mm) and to be Gaussian in transverse direction with $\sigma_r = 1.0$ mm.

2. The beam is assumed to be well focused at the bottom of the blind hole of the collector.

3. Only the collector electrode is considered for the ion-induced secondary electron emission when (say) a beam strikes on it. The total number of back-scattered secondary electrons ($n_{emitted}$) from the collector electrode would be 10 k as we use 1.0 electrons/ion for total ion-SEY and are simulated between energy range of $0 - 100$ eV with peaked at $MPE = 7.5$ eV.

4. A fraction of these electrons are captured ($n_{captured}$) by the blind hole depending on it's depth, while the remaining SEs ($n_{lost} = n_{emitted} - n_{captured}$) either hit on the other parts or leaves the FFC through the beam aperture. The escape probability of the back-scattered SEs is then simply

$$\eta(\%) = \frac{n_{lost}}{n_{emitted}} * 100$$

The escape probability ($\eta$) of the back-scattered SE as function of depth of the blind hole is shown in figure 9 (c). It can be seen that the blind hole in the collector electrode, with its large depth-to-aperture ratio is very efficient and captured almost 80% of the back-scattered SE (geometrical capture). Figure 9 (a) and (b) show the simulated nitrogen and SE trajectories, when depth is equal to 1 mm and 5 mm, respectively.

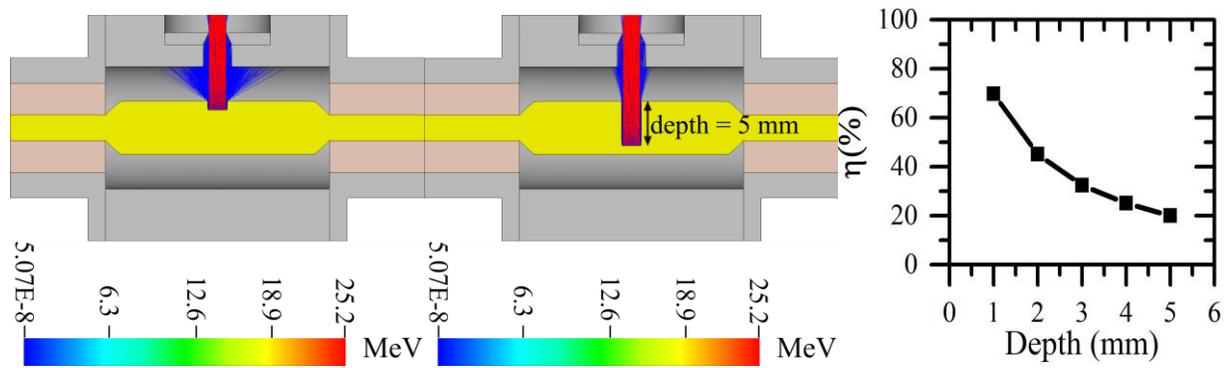

Figure 9. Simulated trajectory of Nitrogen ions (red) and back-scattered SE (blue) when (a) depth = 1mm, (b) depth = 5mm, and (c) the escape probability (h) of the back-scattered SEs as a function of depth of the blind hole.

The escape probability could be reduced further by increasing the depth-to-aperture ratio, but thickness of 1 mm is also required to stop the projectile ions. To further reduce the escape probability of the back-scattered SE, a cone shape was employed in the blind hole of the collector and calculate the escape probability as function of the cone angle ($\psi$). The angle of incidence of the projectile in terms of cone angle is given by $\theta = 90° - \psi$. The calculated escape probability as function of cone angle is presented in figure 10 (b). It should be noted here that the shape of blind hole is purely cone shape when $\psi = 13°$ and it is purely cylindrical when $\psi = 90°$. The minimum escape probability was achieved 3.3% when $\psi = 28°$ (see figure 10 (b)). Figure 10 (a) shows the simulated nitrogen and SE trajectories, when $\psi = 28°$ and $\eta = 3.3\%$.

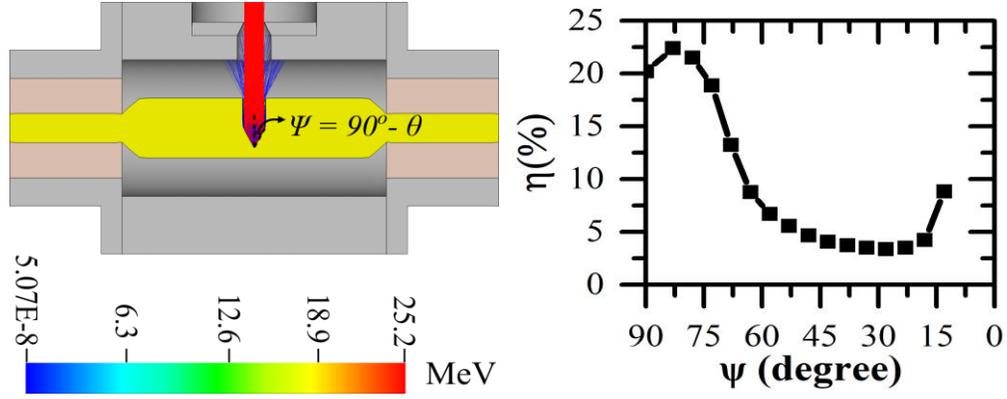

Figure 10. (a) Simulated Nitrogen ions (red) and back-scattered SE (blue) trajectories when cone angle $\psi = 23°$ and (b) The escape probability η of the back-scattered SE as function of cone angle ($\psi$) at 5mm depth of the blind hole.

**Bunch length simulations**

The particle-in-cell (PIC) solver of CST Microwave Studio [11] was used to model the charged particle beam interaction with the FFC structure. The PIC Solver calculates the development of fields and particles through time at discrete time samples. A Gaussian shaped longitudinal charge distribution with a beam bunch charge of 1 fC was used as a excitation source. To demonstrate the induced signals at the N-type connectors by propagation of a bunch beam, a short bunch length of $\sigma = 13.5$ ps with tails cut at $4\sigma$ which is approximately 108 ps full width ($8\sigma$) was chosen to demonstrate the transient FFC signal induction process. For such a bunch travelling with velocity corresponding to $\beta = 0.06$, the full width of the bunch is approximately 2 mm which is less than the blind hole depth of (5 mm) in the collector electrode. The boundaries were set to be open at the N-type connectors to avoid reflections of beam generated fields in the simulation volume. All remaining boundaries were set to a zero tangential electric field. The head of the RCFFC was maintained at ground potential and the inner conductor which serves as collector was set at the floating potential. Table 1 summarizes the input beam parameters used for the simulations. The beamlet formed by the aperture of diameter 2 mm in the collimation disk travels through a gap of length $L = 3.88$ mm between the grounded body and the collector electrode, and is absorbed within a blind hole situated inside the collector electrode. Figure 11 shows a beamlet formed corresponding to a beam of bunch length $\sigma = 13.5$ ps. The signals induced by the beam bunch were collected at the voltage and current monitors defined in the model. Figure 12 (a) demonstrates the propagation of a Gaussian pulse. Figure 12 (b) and 12 (c) are show the induced surface current density and the induced voltage signal on one of the N-type connector, respectively as the bunch propagates through the RCFFC. The time (x-axis) in figure 12 (c) is translated by 94 ps to account for the wave traversal from center of the FFC to one of the N-type connector.

Table 1. Input beam parameters for PIC Solver.

| Parameter | Value |
| --- | --- |
| Bunch Charge | 1 fC |
| Longitudinal distribution | Gaussian ($\sigma \in [13.5, 500]$ ps) |
| Transverse distribution | Uniform (6 mm) |
| Ion species | $N^{5+}$ |
| Velocity | Uniform ($\beta = 0.06$) |
| Bunch length cut-off | $4\sigma$ |
| Macro particles | 5 k |

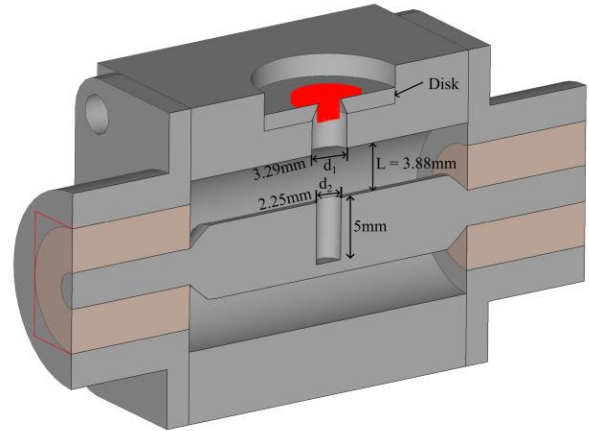

Figure 11. A 3D model of the RCFFC showing beamlet formed by 2 mm diameter hole in the disk for a Gaussian bunch of length $\sigma = 13.5$ ps and $\beta = 0.06$.

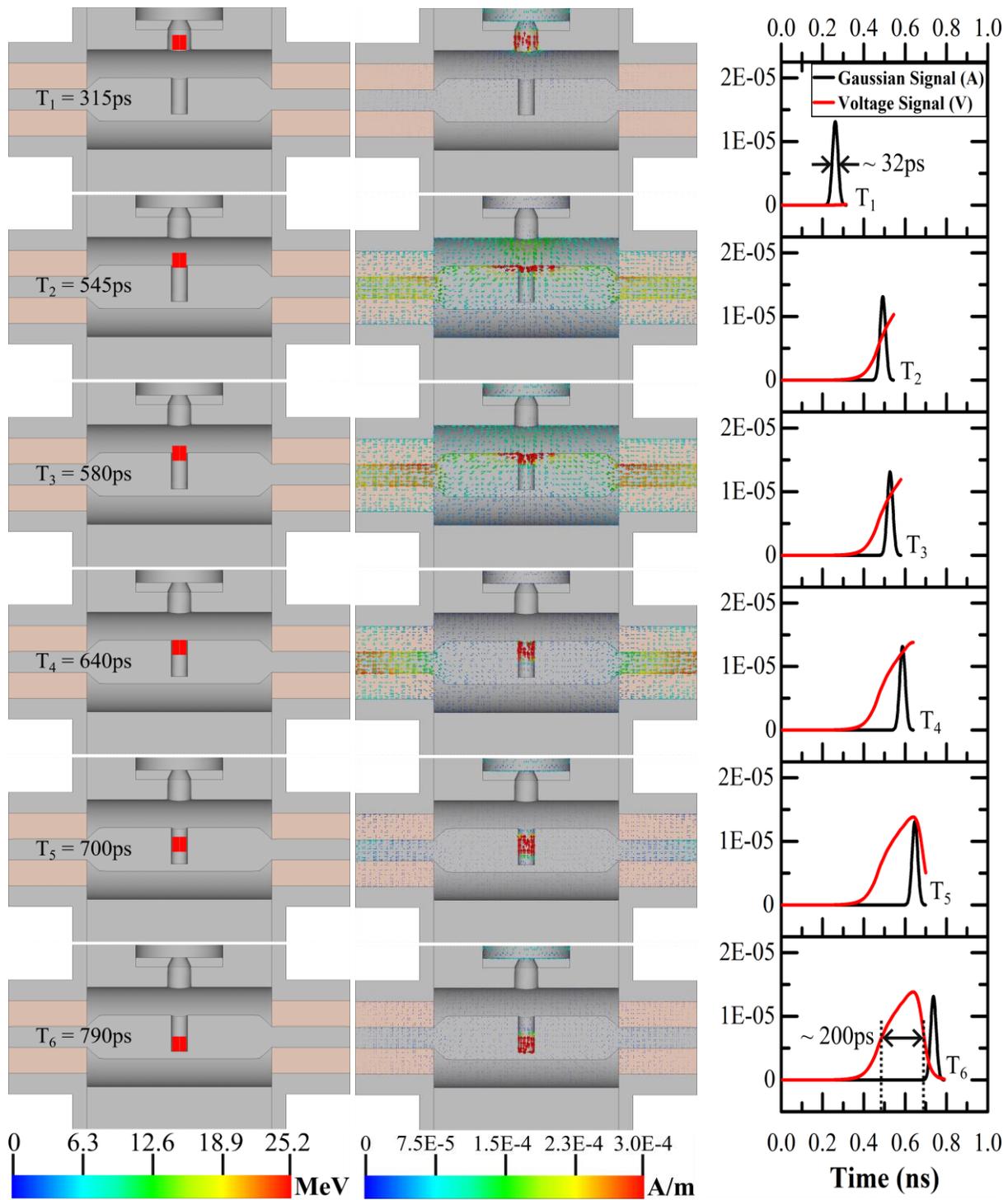

Figure 12. CST Simulation of a Gaussian pulse having bunch length s = 13:5ps with beta=0.06 through the RCFFC showing (a) Propagation of a Gaussian bunch; (b) induced surface current density; and (c) induced voltage signal on one of the N-type connector and a Gaussian pulse.

As we can see from figure 12., the voltage signal starts to induce, once the bunch starts to enter in the gap and rises as the bunch approaches the collector electrode. The induced signal attains the highest value (at time $T_2 = 545$ ps), when the bunch about to enter in the blind hole of the collector electrode. At this point ($T_2 = 545$ ps), the maximum density of the opposite charges (electrons) induces on the collector electrode (see figure 12 (b)) and the same density of the electrons travels with the bunch when bunch starts to enter into the blind hole. As a result, the signal starts to reduce and reaches zero at time $T_5 = 700$ ps (see figure 12 (c)) before absorbing the bunch into the collector electrode where negative charges neutralised with positive charges of the bunch. Also we can see from figure 12 (b), after time $T_4 = 640$ ps, there is no flow of current from center of the FFC to the N-types connectors except the earlier induced current. That means the induced signal should be reduced to zero once the bunch completely entered into the hole. But it reduced to zero at time $T_5 = 700$ ps and the full width at half maximum (FWHM = 2.3548 $\sigma$) of the induced signal is approximately 200 ps as shown in figure 12 (c). Therefore, the bunch length of the induced signal will be $\sigma_c = 85$ ps which is larger than the bunch length of the input beam ($\sigma_b = 13.5$ ps).

**The effect of hole size and the finite gap**

The large size of the holes in the ground electrode and in the blind hole can modify the measured signal from the actual shape of the original bunch. It can be seen from figure 11, the dimensions of both the diameters ($d_1$ and $d_2$) are comparable with the gap length (L). Because of the large size of these diameters, the electric field of the bunch causes a motion of charges in the collector electrode before the entry of the bunch in the gap. This effect is demonstrated in figure 13. Figure 13 (a) shows the propagation of a Gaussian bunch through the gap when center of the bunch entered in the gap ($t_1 = 370$ ps) and when center of the bunch leaving the gap ($t_2 = 580$ ps).

Figure 13 (b) shows the induced voltage signal and its derivative for two different sets of d1 and d2.

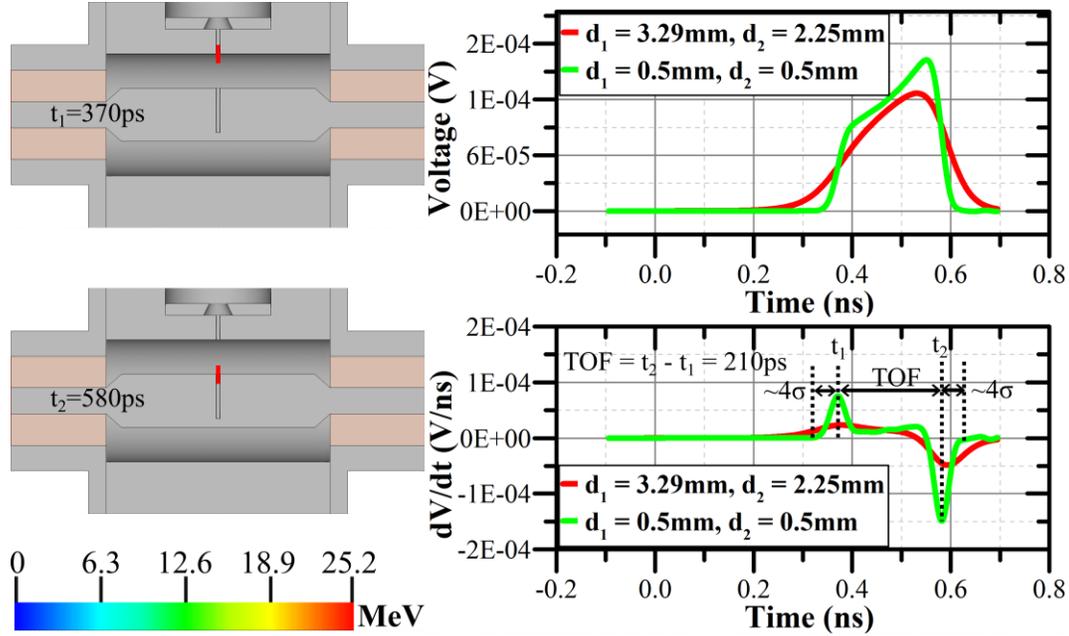

Figure 13. CST Simulation of a Gaussian pulse with bunch length s = 13:5ps with beta=0.06 through the RCFFC showing (a) Propagation of a Gaussian pulse; (b) Effect of d1 and d2 on the induced voltage signal and its derivative.

When d1 and d2 are reduced from their optimized values (see figure 11) to 0.5mm, the difference between the maxima and minima of the derivative of induced voltage signal becomes exactly equal to the time of flight of the particles (TOF = 210 ps) in the gap. The bunch length of the induced voltage signal is also reduced when the hole sizes are reduced but it can't be reduced further as the gap length of 3.88mm is required for 50 $\Omega$ impedance. Therefore, there will be a widening in the measured signal due to the finite gap and the large size of the holes.

The effect of the finite gap on the induced signal is also explained in a theoretical model, where the hole size has taken very small compare to the gap length and assumed no effect on the induced signal. The measured bunch length can be calculated as follows:

$$\sigma_c^2 = \sigma_b^2 + \sigma_{gap}^2 \qquad (8)$$

Where $\sigma_{gap} = \kappa.\tau$ is the bunch length due to the gap. $\tau = L/v$ is the time of flight of particles and $v$ is the particle velocity. $\kappa$ is a numerical constant determined by the FFC geometry and estimated to be $\kappa \approx 0.29$ [30]. The relative error in measuring the bunch length can be calculated as follows [30]:

$$\frac{\sigma_c}{\sigma_b} - 1 = \sqrt{1 + \left(\frac{\sigma_{gap}}{\sigma_b}\right)^2} - 1 \qquad (9)$$

Considering the input beam parameters $\sigma_b = 13.5$ ps, $v = 18.57$ mm/ns for $\beta = 0.06$, the bunch length due to gap will be $\sigma_{gap} = 60$ ps and the measured bunch length will be $\sigma_c \approx 62$ ps using equation 8. The bunch lengths of the induced signal (see figure 13 (b)) are 73 ps and 83 ps respectively, for the case 1 ($d_1 = d_2 = 0.5$ mm) and case 2 (optimized values of $d_1 = 3.29$ mm and $d_2 = 2.25$ mm). The relative error in measuring the bunch length is 359 % at $\sigma_b = 13.5$ ps. Similarly, the relative errors in measuring the bunch length of the simulated signal are 440% and 515% for case1 and case 2, respectively at $\sigma_b = 13.5$ ps. A comparison of theoretical and simulated relative errors in measuring the various input bunch lengths are summarized in table 2 and the corresponding simulated voltage signals are shown in figure 14. Figure 14 (a) and 14 (b) are illustrating the induced voltage signal in the time domain for case 1 and case 2, respectively as function of input bunch length $\sigma_b$ for $\beta = 0.06$.

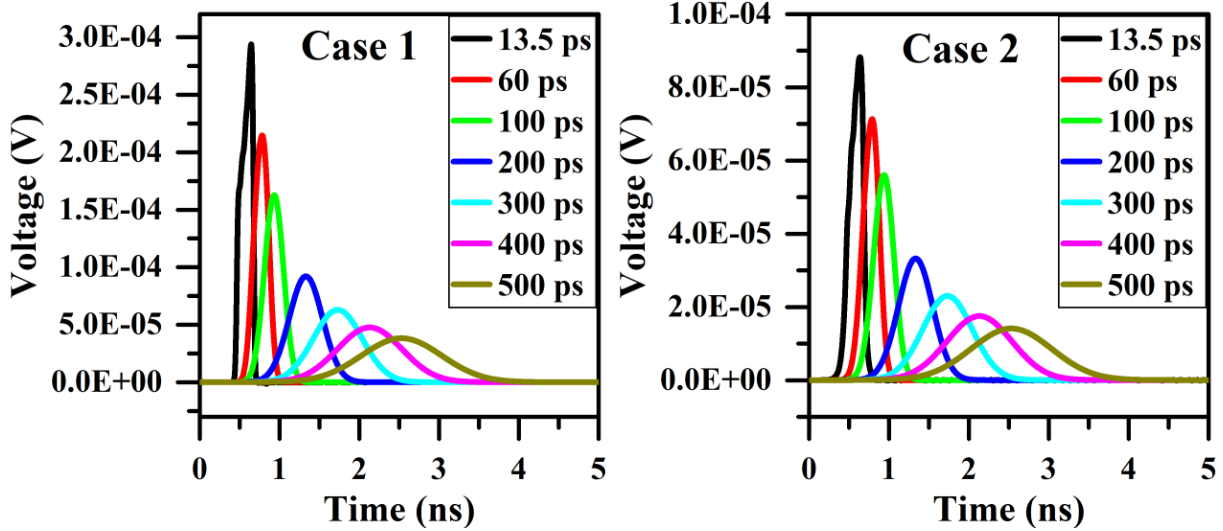

Figure 14. Simulated induced voltage signal in the time domain as a function of the input bunch length $\sigma_b$ having β = 0.06 and 1fC bunch charge for (a) Case 1 ($d_1 = d_2 = 0.5$ mm), and (b) Case 2 ($d_1 = 3.29$ mm, $d_2 = 2.25$ mm).

**Table 2. Comparison of theoretical and simulated relative error in measuring $\sigma_b$ for various input bunch lengths.**

| Input Pulse | Finite gap | Theoretical Model | | Simulated Model | | | |
| --- | --- | --- | --- | --- | --- | --- | --- |
| | | | | case 1 | | case 2 | |
| $\sigma_b$(ps) | $\sigma_{gap}$(ps) | $\sigma_c$(ps) | % | $\sigma_s$(ps) | % | $\sigma_s$(ps) | % |
| 13.5 | 60 | 62 | 359 | 73 | 440 | 83 | 515 |
| 60 | 60 | 85 | 42 | 89 | 49 | 100 | 66 |
| 100 | 60 | 117 | 17 | 118 | 18.4 | 127 | 27 |
| 200 | 60 | 209 | 4.5 | 209 | 4.5 | 215 | 7 |
| 300 | 60 | 306 | 2 | 306 | 2 | 310 | 3 |
| 400 | 60 | 404 | 1.1 | 404 | 1.1 | 407 | 1.9 |
| 500 | 60 | 503 | 0.7 | 503 | 0.7 | 506 | 1.2 |

From table 2, one can see that the relative error in measuring the bunch length for the theoretical and case 1 of the simulated model are almost same for input bunch length higher than 60 ps, however, there is a difference for case 2. This is because of the hole size which also creates a widening in the induced signal in addition of the gap length. The bunch length of the induced signal is mainly defined by the gap length and the diameters of the holes in the RCFFC for the input bunch length lower than 60 ps. As a result, the differences in the relative error of case 1 and case 2 are minimized as the input bunch length increases. Considering equation 9 with the input beam of having velocity $v = 18.57$ mm/ns for $\beta = 0.06$, the relative error in measuring $\sigma_b$ is < 42 % at $\sigma_b > 60$ ps. Similarly, the relative error in measuring $\sigma_b$ in the simulation for case 2 is 66 % (see table 2.). Thus, the current design of the RCFFC is only suitable to measure the pulses of bunch width above 200 ps whereas the relative error is lower than 10 % as can be seen in the table 2.

There can be a substantial difference between the bunch length of the input beam and that of the induced signal just on the FFC structural parameters, e.g. the effect of hole size and the finite gap in the FFC. Other effects like the secondary electron emission (SEE), the dispersion and reflection in the connecting cables, and bandwidth of the fast oscilloscope come on top of this effect. The SEE affects the induced signal as discussed in the next section.

**The effect of Secondary Electron Emission**

To consider the effect of the secondary electron emission (SEE) in the bunch length simulation, we use the Vaughan Model [31] for electron induced SEE which is also a built-in model in CST Microwave Studio and the theoretical formulas as mentioned in equation 5 and 6 for ion-induced SEE. The maximum energy of primary and secondary electrons were 100 eV. The secondaries released by the ion-induced SEE were considered as a source of the primary electrons in the simulation for the electron-induced SEE. The values of ion-SEY (secondary electron yield) and electron-SEY are used 1.0 electron/ion for energy 25.2 MeV (considering the case for N5+) and 2.1 electrons/electron (default value) for electron energy 100 eV. Equation 6 provides the energy spectrum of the SEs produced by the primary ions and electrons, and $MPE = 7.5$ eV was chosen as the energy at which the most SEs would be emitted. These energy spectra which were used in the simulation are shown in figure 15. With the exception of the N-type connector, which uses Teflon, all RCFFC components were taken into account for SEE and used copper material. The effect of SEE on the induced voltage signal as a function of the input bunch length ($\sigma_b$) for $N^{5+}$ ions of having energy 1.8 MeV/A or $\beta = 0.06$ is shown in figure 16 (a). The induced voltage due to effect of the SEE is low compared to the amplitude of the direct ion induced voltage signal. This is because of the large depth to aperture ratio (see figure 9 (c)), where the escape probability of SEE was significantly reduced from 70% to 20% by employing a blind hole of depth 5 mm in the collector electrode. One can see that the spectrum of the SEE are completely separated from the induced signal for input bunch length $\sigma_b < 200$ ps, but alter the shape of the induced signal for $\sigma_b > 200$ ps.

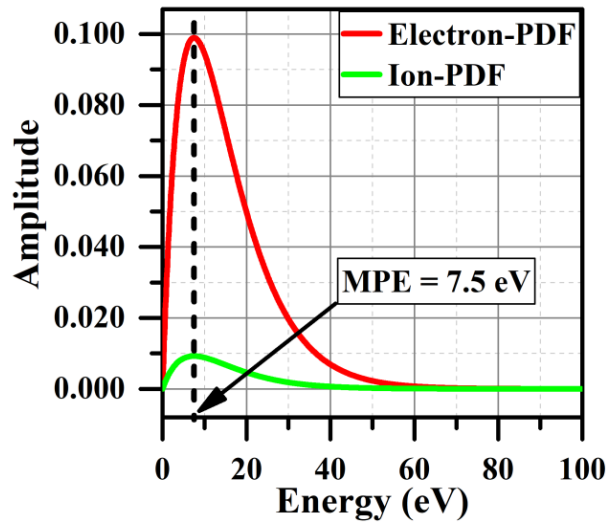

Figure 15. Secondary electron energy spectra of ion and electron-induced SEE for the input parameters Ion-SEY $= 1.0$ electrons/ion, Electron-SEY $= 2.1$ electrons/electron, MPE $= 7.5$ eV, and $\theta = 0°$.

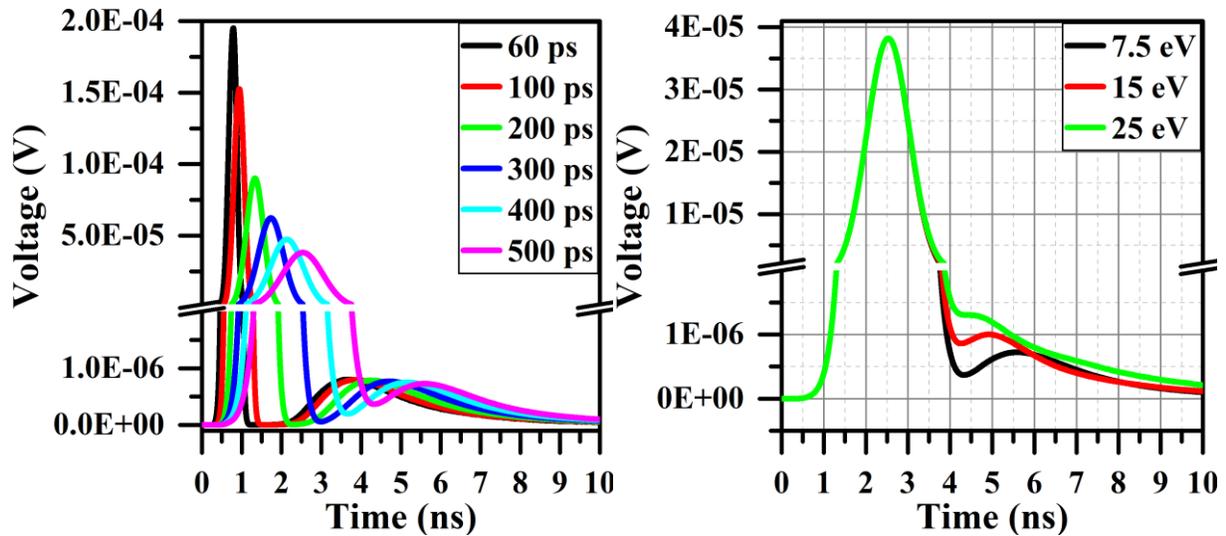

Figure 16. Simulated induced voltage signal in the time domain for bunch having b = 0.06 and 1fC bunch charge (a) As a function of the input bunch length sb for MPE = 7.5eV, and (b) As a function of MPE for the input bunch length $\sigma_b$ = 500 ps.

The peak position of the SEE spectrum in the time domain can be estimated. Considering the induced voltage signal for input bunch length 500 ps, $v = 18.57$ mm/ns and the peak position of the induced signal is approximately 2.5 ns (see figure 16 (a)) which corresponds to the center of the bunch about to enter in the blind hole. Such a peak was recently seen in an FFC measurement. The maximum number of the SE emitted when the center of the bunch hits the bottom of the blind hole and the time taken by center of the bunch to reach at the bottom of the blind hole is 0.27 ns. The default value 7.5 eV or 1.62 mm/ns for the MPE was used and these electrons will take 3 ns time to enter in the gap after emitted from bottom of the blind hole. Therefore, the peak position of the SEE spectrum in time domain will be 5.8 ns for input bunch length of 500 ps, which can also be seen in figure 16 (a). The effect of the MPE on the peak position of the SEE spectrum and the induced signal can be seen in the figure 16 (b), for input bunch length 500 ps. As the mean value of MPE increases, the peak position of the SEE spectrum shifted towards the induced signal. At the higher MPE, the peak of the SEE spectrum could be completely merged into the induced signal and can significantly change the actual shape. We have also analyzed the effect of the static (dc) voltage on the collector to suppress the SEE. To suppress the SEE, the head of the RCFFC was maintained at ground potential and the potential on the collector was varied from $-25$ V to 25 V. The effect of the dc biasing on the SEE spectrum can seen in the figure 17. The negative voltage on the the collector electrode repels the secondary electrons from the collector electrode, as a result the amplitude of SEE spectrum increases with reducing the voltage. The positive voltage on the collector electrode, attracts the secondary electrons towards the collector, and keeps the secondary electrons of energy below the applied dc voltage inside the blind hole. As a result, the amplitude of the SEE spectrum reduces with increasing the positive dc voltage. But there is a negative peak of the SEE, and amplitude of this peak increases with increasing the positive voltage. This may be due to the SE of having energy more than 25 eV escaped from the blind hole and hits on the disk which further starts the SEE. Due to the positive voltage on the collector electrode, these SE would be accelerates towards the collector electrode and a negative signal would be produced depending the dc voltage on the collector electrode (see figure 17).

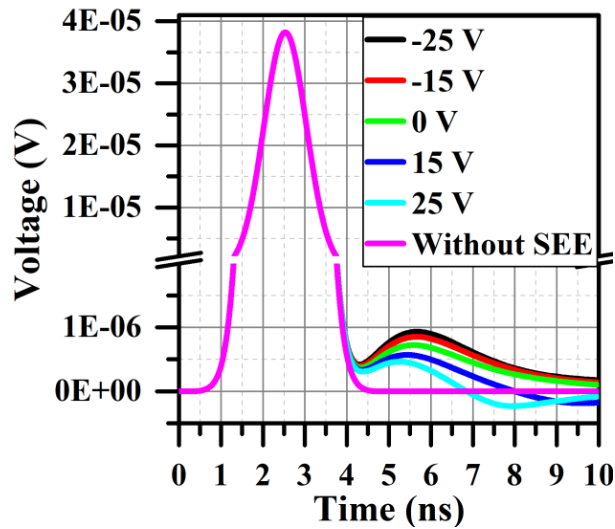

Figure 17. Simulated induced voltage signal with and without SEE in the time domain as a function of the dc biasing for input parameters $\sigma_b = 500$ ps, $\beta = 0.06$, and MPE $= 7.5$ eV.

**Thermal Simulations**

The dielectric constant of Teflon is found to decrease with the temperature [32], therefore, if the temperature of Teflon rises due to the thermal load produced by the projectiles, the characteristic impedance of the RCFFC would accordingly change. Therefore, a detailed thermal analysis of the RCFFC structure needs to be looked into detail. The thermal load produced by the projectiles depends on the beam energies and intensities. The proposed design is generally intended for measuring bunch lengths from lower to heavier ion beams having A/q <= 6, having energies ranging from 8 keV/A to 1.8 MeV/A and with beam intensities in the order of few particle µA. Therefore the maximum thermal load can be calculated by the following equation

$$P = VI \qquad (10)$$

where P is the thermal load in Watts, and I is the beam intensity in µA. V is the total accelerating voltage of the beam in million Volts and can be calculated by the following equation

$$V = \frac{E_{max} * A/q}{e} \qquad (11)$$

where $E_{max}$ is the maximum energy in MeV/A. A, q, e are the mass in atomic mass units (a.m.u), charge state, and electronic charge of electron, respectively. Thus, putting equation 11 into 10 and using appropriate numerical values, the maximum thermal load will not exceed 10.8 W for a beam of having A/q = 6 and beam intensity of 1 µA. Considering a DC beam intensity of 1 µA for 10 minutes, the maximum temperature of the collector electrode for different thermal load are calculated using COMSOL Multiphysics [33] and is shown in figure 18.

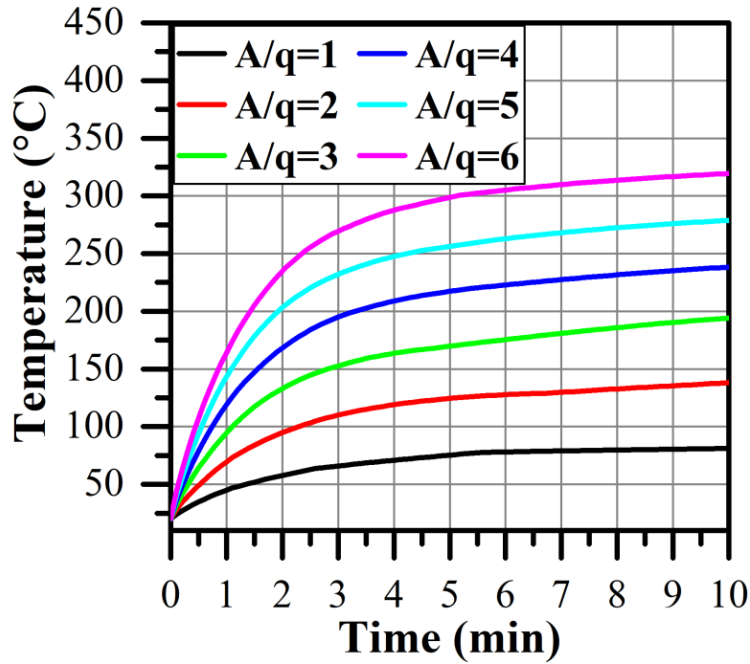

Figure 18. Evolution of temperature of the collector electrode for varying thermal loads with beam intensity of 1 A.

It can be seen that the temperature of the collector electrode crossed the melting point ($327^O$ C) of Teflon in 10 minutes for a DC beam. Therefore, water cooling is essential for DC beam. It is to be noted that the actual deposited power by the pulsed beam would be less compared to the case of a DC beam.

At the High Current Injector [2,3], the pulsed beam ( bunch length of the order of few ns) is generated by a Multi-Harmonic Buncher [34], which operates at a fundamental frequency of 12.125 MHz. Considering a bunch length of 4 ns in the worst possible case, the duty cycle of the pulse would be 4.85 %. In the simulation, the step size should less than the width of the pulse otherwise the thermal load calculation would not be activated. To simulate the RCFFC for thermal analysis using the pulse of width 4 ns with duty cycle 4.85 % and step size of 1 ns, the total number of steps between 0 to 10 minutes would be 6E11 . It is impossible to handle such large number of steps for a normal computer (16 GHz RAM with 4 cores). Therefore, we use the frequency

12.125 mHz instead of 12.125 MHz to reduce the total number of steps in the simulation. The pulse width now becomes 4 s and the total number of steps would be 600 from 0 to 10 minutes. The rectangular pulse of width 4 s and duty cycle 4.85 % for a pulsed beam of A/q = 6 is shown in figure 19.

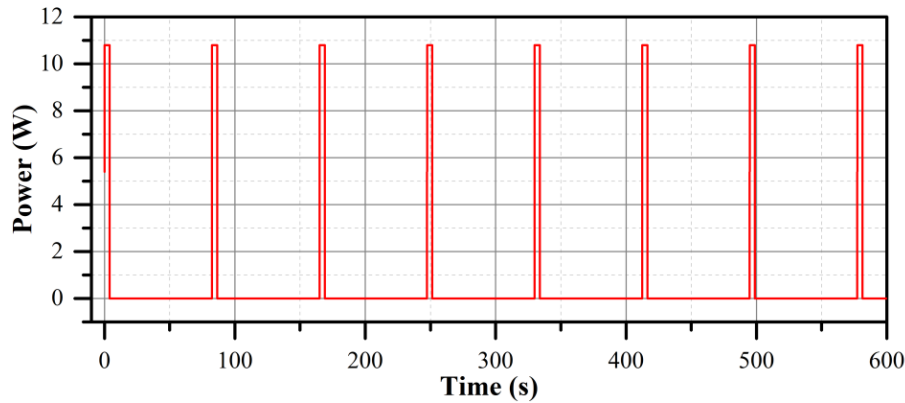

Figure 19. Rectangular pulse with beam parameters: A/q = 6, beam intensity of 1 µA, pulse width = 4 s, duty cycle = 4.85 %.

The evolution of the temperature of the collector electrode and average temperature of the RCFFC for the pulsed beam is shown in figure 20 (a). The temperature distribution of the RCFFC for pulsed beam is shown in figure 20 (b). After a 10 minute run of the pulsed beam, we can see that the maximum temperature of the collector electrode and the average temperature of the RCFFC have been reached, respectively at 50º C and 26º C. These temperature values are much below the melting point of Teflon. Therefore, no water cooling is required for the RCFFC in the case of bunched beams.

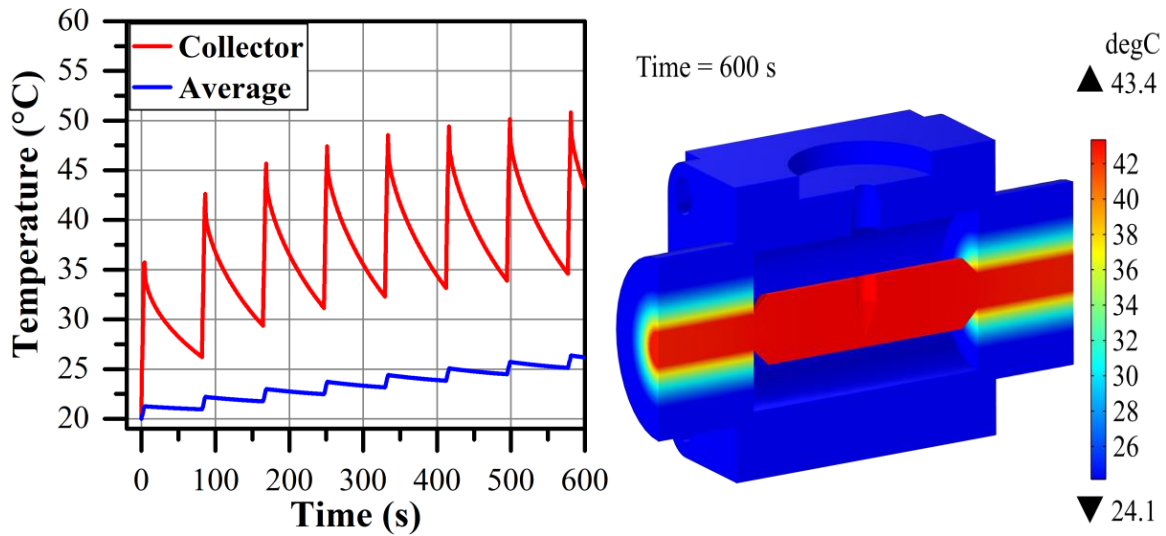

Figure 20. Thermal simulation of RCFCC with a pulsed beam of parameters: A/q = 6, beam intensity = 1µA, pulse width = 4s, and duty cycle = 4.85 % showing (a) Evolution of temperature of collector electrode and average temperature, and (b) 3D temperature distribution

**Measurement of the scattering parameters of the first prototype**

The first prototype was fabricated at IUAC and lab tests were carried out at GSI to determine the scattering parameters in order to validate the EM design. Figures 21 shows the reflection parameters S11 and S22 while Figure 22 show the S21 parameters measured until 9 GHz. The measured values are comparable with the expected parameters with 0.1 mm fabrication errors discussed in Fig. 7 thus validating the basic design. We assign most of the deviations of the fabricated FFC from the EM design to the imperfect interfacing to the N connectors.

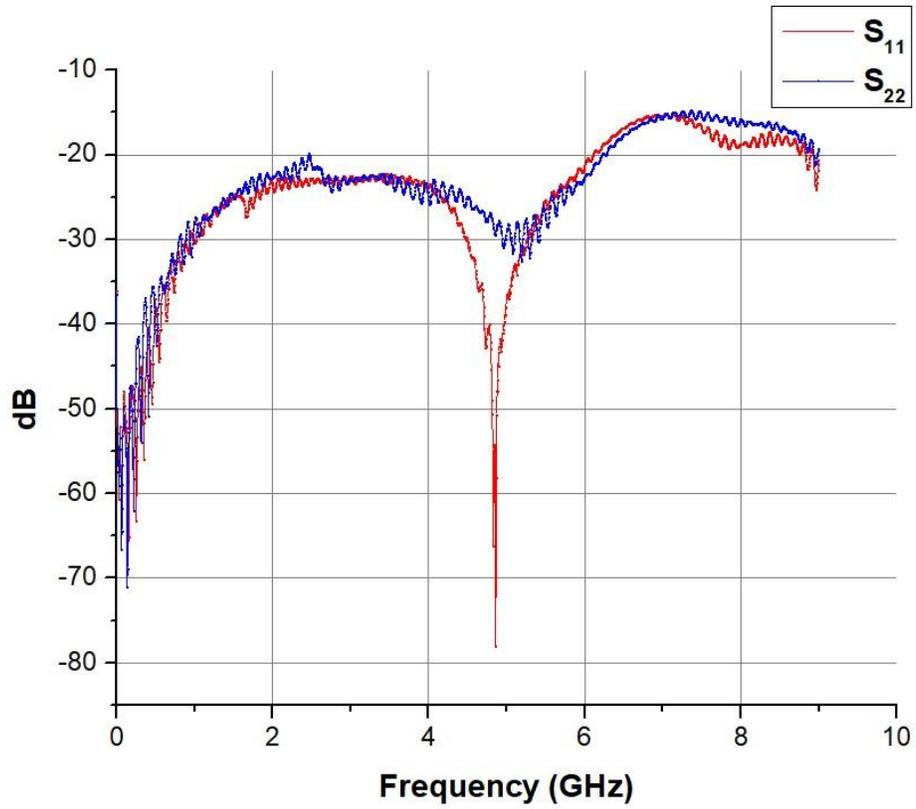

Figure 21. Measured $S_{11}$ and $S_{22}$ parameters of the first prototype. Large difference between S11 and S22 suggest improper alignment and soldering of the central conductor to the N connectors.

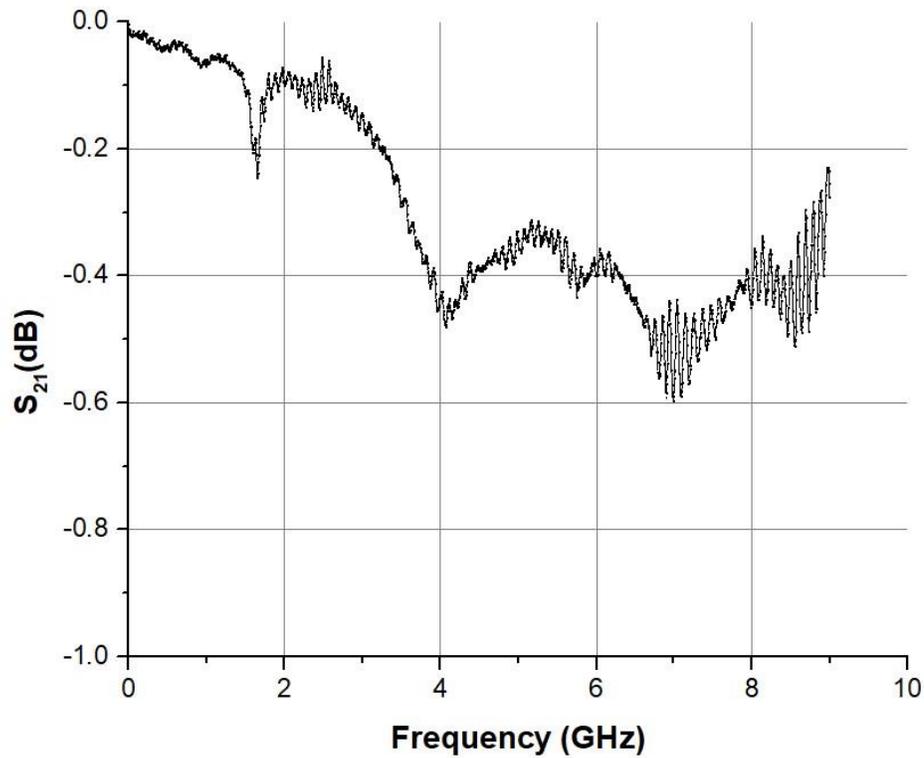

Figure 22. Measured S$_{21}$ parameters of the first prototype

**Conclusion**

The radially coupled co-axial FFC (RCFFC) design for high intensity proton beams is modified for low intensity ion beams. The central conductor and the hole for the beamlet entry were widened while maintaining the characteristic impedance of the full structure close to 50 Ω. A transition method without curved structures was adopted for the transition from the N-type connector to the cup region to achieve uniform impedance and low reflection. Adaptations in the shape of the blind hole surface in the central conductor was made to reduce the number of secondary electrons exiting the blind hole. Signal induction procedure and its dependence on the FFC design parameters, like entry hole size and gap length was discussed in detail. The effect of secondary electron emission on the induced signal is studied and potential remedy is proposed. Heating of the FFC for dc and bunched beams is simulated and it is shown that there is no need for water cooling for bunched beams planned at HCI in IUAC. Finally, network analyzer measurements of the first prototype is presented which confirms the EM design of the IUAC RCFFC.

**Acknowledgement**

One of the authors, R. Singh (GSI, Darmstadt) was invited as a VAJRA (Visiting Advanced Joint Research) Adjoint faculty to IUAC, New Delhi during the course of this study. R.Singh and G. Rodrigues would like to acknowledge the help and support received from the Department of Science and Technology (DST) related to the VAJRA project (VJR/2018/000115).

**References**

[1] Carneiro, J.-P., et al. "Longitudinal Beam Dynamics Studies at the Pip-II Injector Test Facility." International Journal of Modern Physics A, vol. 34, no. 36, 2019, p.1942013.

[2] Roy, A. (1999). Accelerator development at the Nuclear Science Centre. Current Science, 76(2), 149–153. http://www.jstor.org/stable/24101229

[3] D Kanjilal, C P Safvan, G O Rodrigues, P Kumar, U K Rao, A Mandal, A. Roy, High current injector at Nuclear Science Centre, in: Proc. APAC Korea (2004) 73.

[4] P. Strehl, Beam Instrumentation and Diagnostics (Springer-Verlag Berlin Heidelberg, 2006).

[5] H. Chuaqui, M. Favre, E. Wyndham, and L. Arroyo, Rev. Sci. Instrum. 60, 141(1989).

[6] M. Bellato, A. Dainelli, and M. Poggi, Nucl. Instrum. Methods Phys. Res., Sect. A 382, 118 (1996).

[7] W. R. Rawnsley, R. E. Laxdal, L. Root, and G. H. Mackenzie, AIP Conf. Proc. 546, 547 (2000).

[8] Van der Walt, P. W. (n.d.). "A novel matched conical line to coaxial line transition". Proceedings of the 1998 South African Symposium on Communications and Signal Processing-COMSIG '98 (Cat. No. 98EX214). https://doi.org/10.1109/comsig.1998.736998

[9] D.M. Pozar, "Microwave engineering", 2nd edition, 1998 John-Wiley & Sons.

[10] Park, H. H. (2021). Design of compact transition from conical to coaxial transmission lines with a low return loss. International Journal of Electronics, 108(8), 1426–1438.

[11] See www.3ds.com/cst-studio-suite/ for details of simulation code CST microwave studio.

[12] S. Y. Liao, Microwave Devices and Circuits (Prentice Hall, NJ, 1985).

[13] TDR (Time Domain Reflectometry), https://space.mit.edu/RADIO/CST_online/mergedProjects/3D/common_preloadedmacro/common_preloadedmacro_1d_tdr.htm.

[14] J.D. Thomas, G.S. Hodges, D.G. Seely, N.A. Moroz, and T.J. Kvale, Nucl. Instrum. Meth. Phys. Res. A 536, 11–23 (2005).

[15] Ebrahimibasabi, E., and Feghhi, S. (2015). Design and construction of a secondary electron suppressed Faraday Cup for measurement of beam current in an electrostatics proton accelerator. International Journal of Mass Spectrometry, 386, 1–5.

[16] Cantero, E., Sosa, A., Andreazza, W., Bravin, E., Lanaia, D., Voulot, D., and Welsch, C. (2016). Design of a compact Faraday cup for low energy, low intensity ion beams. Nuclear Instruments and Methods in Physics Research Section A: Accelerators, Spectrometers, Detectors and Associated Equipment, 807, 86–93.

[17] Mathew, J. V., and Bajaj, A. (2020). An improved strip-line fast Faraday cup for beam bunch measurements. Review of Scientific Instruments, 91(11), 113305. https://doi.org/10.1063/5.0025457

[18] G.F. Knoll Radiation Detection and Measurements, vol. 2, John Willey and Sons, Inc., New York (2000), p. 30.

[19] D. Hasselkamp, H. Rothard, K.O. Groeneveld, J. Kemmler, P. Varga, H. Winter Particle Induced Electron Emission 2, vol. 4, Springer-Verlag, Berlin/Heidelberg (1992), p. 25

[20] J.B. Marion Classical Dynamics of Particles and Systems (second ed.), Academic Press, New York (1970).

[21] P. Sigmund, S. Tougaard, Springer Ser. Chem. Phys. 17 (1981) 2.

[22] R.A. Baragiola, E.V. Alonso, A. Oliva Florio Electron emission from clean metal surfaces induced by low-energy light ions Phys Rev B, 19(1) (1979,), pp. 121-129.

[23] A. Koyama, T. Shikata, H. Sakairi Secondary electron emission from Al, Cu, Ag and Au metal targets under proton bombardment Jpn J Appl Phys, 20 (1) (1981), pp. 65-70.

[24] D. Hasselkamp, K.G. Lang, A. Scharmann, N. Stiller Ion induced electron emission from metal surfaces Nucl Instrum Meth, 180 (2–3) (1981), pp. 349-356.

[25] Murdock JW, Miller GH. Secondary electron emission due to positive ion bombardment. Ames Laboratory ISC Technical Reports. 106; 1955.

[26] O. Benka, A. Schinner, T. Fink, M. Pfaffenlehner Electron-emission yield of Al, Cu, and Au for the impact of swift bare light ions Phys Rev A, 52 (5) (1995), pp. 3959-3965.

[27] Haque, A. F., Haque, M., Sultana, S., Patoary, M. A. R., Hossain, M. S., Maaza, M., and Uddin, M. A. (2019). Proton-induced secondary electron emission from elemental solids over the energy domain 1 keV–1000 MeV. Results in Physics, 15, 102519. https://doi.org/10.1016/j.rinp.2019.102519

[28] J. Ziegler, Particle Interactions with Matter. Available from: http://www.srim.org, 2006.

[29] see the link https://www.nist.gov/pml/stopping-power-range-tables-electrons-protons-and-helium-ions

[30] A. Shemyakin, arXiv:1612.09539.

[31] J. R. Vaughan. Secondary Emission Formulas. IEEE Transactions on Electron Devices, 40(4), 1993.

[32] Ehrlich, P., Amborski, L. E., and Burton, R. L. (1953). Dielectric properties of Teflon from room temperature to 314°C and from frequencies of 102to 105c/s. 1953 Conference on Electrical Insulation. https://doi.org/10.1109/eic.1953.7508671

[33] COMSOL Multiphysics® v. 5.6. www.comsol.com. COMSOL AB, Stockholm, Sweden.

[34] Sarkar, A., Ghosh, S., Barua, P., Joshi, R., Ahuja, R., Rao, S., Krishnan, S. A., Malyadri, A. J., Kumar, R., Gargari, S., Chopra, S., Kanjilal, D., Datta, S. K., Roy, A., Bhowmik, R. K., Tilbrook, I. R., & Clifft, B. E. (1999). Single-gap multi-harmonic buncher for NSC pelletron. The Eighth International Conference on Heavy-Ion Accelerator Technology. https://doi.org/10.1063/1.58948